\documentstyle[12pt]{article}
\begin{document}
\begin{center}
\begin{flushright}
{}\hfill
MRI-PHY/P991236
\end{flushright}
\vspace{10mm}
{\Large{\bf{Nonsingular static global string}}}\\
\vspace{5mm}
A.A.Sen{\footnote{anjan@mri.ernet.in}}\\
Mehta Research Institute\\
Chhatnag Road, Jhusi,\\
Allahabad,211019, India\\
\vspace{5mm}
and\\
\vspace{5mm}
N.Banerjee{\footnote{narayan@juphys.ernet.in}} \\
Relativity and Cosmology Research Center\\
Department of Physics\\
Jadavpur University\\
Calcutta 700032\\
India.
\end{center}
\vspace{10mm}
{\centerline{\underline{Abstract}}}
{\small{
A new solution for the spacetime outside the core of a $U(1)$ static
global string has been presented which is nonsingular. This is the first
example of a nonsingular spacetime around a static global string.}}

\vspace{10mm}
It is believed that early universe has undergone a number of phase
transitions
as it cooled down from the hot initial phase. One of the immediate
consequences of these phase transitions is the formation of defects or
mismatches in the orientation  of the Higgs field in causally disconnected
regions~\cite{Kibble}. One of these defects, cosmic string, is particularly 
interesting
as it is capable of producing observational effects and also may play an
important role in the large scale structure formation of the universe.The
gravitational field of a static infintely long gauge string, produced
by the breaking of a $U(1)$ local symmtery, has been studied in great
detail~\cite{Local}. It has been shown in all of these investigations
that the spacetime  around a infinitely long local gauge string is
Minkowskian minus a wedge. However for global strings which arise due
to the breaking of global symmetry, the exact gravitational field  is
not that simple. Global strings are such that their energy extends to
regions far beyond the central core, the energy density 
being proportional to $r^{-2}$, which leads to a logarithmically
divergent mass per unit length. For a global string, Harari and
Sikivie~\cite{HS} presented a solution of the linearised Einstein's field
equations neglecting the radial variation of the scalar field
outside the core of the string. That the approach was not self
consistent became evident as there is a physical singularity at a
finite distance from the string core. Cohen and Kaplan~\cite{CK}
considered the exact nonlinear field equations.  In their solution, 
there was a unexplained
singularity on the symmetry axis while the string appeared as a
singular cylindrical surface of finite radius. The present authors  
have considered~\cite{Ban} the
Einstein's field equations outside the core of a static global
string, in a coordinate system different from that considered by
Cohen and Kaplan. It has been shown that our solution
reduces to the Cohen and Kaplan solution under a coordinate
transformation.
That an undesirable singularity is unavoidable for the static global
string was later pointed out by Gregory~\cite{Greg1} who ascribed it to
a pecularity of the energy momentum tensor rather than to  the slow
fall off of the energy density with radial distance. 
Gibbons et.al showed that the global string cannot admit any asymtotically well
behaved solution as the deficit angle produced by the string diverges
logarithmically for a large distance~\cite{Gib}. Later on,
Gregory~\cite{Greg2} showed that one can get a nonsingular solution
for the spacetime outside the core of a global string only if one includes
a time dependence along the symmtery axis. It should be mentioned
that in all of these work, it has been assumed that the spacetime
admits a Lorentz boost along the symmetry axis. In the present work,
we have been able to find a nonsingular solution for the spacetime
outside the core of a global string. For achieving such a solution,
we have relaxed the condition of a  Lorentz boost being admitted along the
symmetry axis of the string.

To describe the spacetime geometry due to a infinitely long static
cosmic string, the line element  is taken to be the general static
cylindrically symmetric one given by
$$
ds^{2}= e^{2(K-U)}(dt^{2}-dr^{2})-e^{2U}dz^{2}-e^{-2U}W^{2}d\theta^{2},
\eqno{(1)}
$$
where $K, U, W$ are functions of $r$ alone. For a global string, the
energy momentum tensor components are calculated from the action
density for a complex scalar field $\psi$ along with a Mexican hat potential:
$$
L={1\over{2}}g^{\mu\nu}\psi^{\star}_{,\mu}\psi_{,\nu}-{\lambda\over{4}}
(\psi^{\star}\psi-v^{2})^{2},
\eqno{(2)}
$$
where $\lambda$ and $v$ are constant and
$\delta=(v\sqrt{\lambda})^{-1}$ is a measure of the core radius of the
string. It has been shown that ~\cite{Greg1} the field configuaration
can be choosen as 
$$
\psi(r) = v f(r) e^{i\theta}
\eqno{(3)}
$$
in cylindrical field coordinates. The usual boundary condition on
$f(r)$ is $f(0)=0$ and $f(r)\rightarrow 1$ as $r\rightarrow \delta$.
As we are interested in spacetime outside the core of the string, for
our purpose
$$
f(r) = 1 , f^{\prime}(r) = 0
\eqno{(4)}
$$
is a good  approximation. The nonzero components of the energy
momentum tensor outside the core of the string now become
$$
T^{t}_{t}=T^{r}_{r}=T^{z}_{z}=-T^{\theta}_{\theta}=
{v^{2}\over{2}}{e^{2U}\over{W^{2}}}.
\eqno{(5)}
$$
The Einstein's equations, $G^{\mu}_{\nu} = 8\pi T^{\mu}_{\nu}$, are
$$
-{W^{\prime\prime}\over{W}}+{K^{\prime}W^{\prime}\over{W}}-U^{\prime2}
=-{4\pi v^{2}\over{W^{2}}}e^{2K},
\eqno{(6a)}
$$
$$
-{K^{\prime}W^{\prime}\over{W}}+U^{\prime2}
=-{4\pi v^{2}\over{W^{2}}}e^{2K},
\eqno{(6b)}
$$
$$
-K^{\prime\prime}-U^{\prime2}
={4\pi v^{2}\over{W^{2}}}e^{2K},
\eqno{(6c)}
$$
$$
-{W^{\prime\prime}\over{W}}+2U^{\prime\prime}+2U^{\prime}{W^{\prime}\over{W}}
-K^{\prime\prime}-U^{\prime2}=-{4\pi v^{2}\over{W^{2}}}e^{2K}.
\eqno{(6d)}
$$
By adding equations (6a) and (6b) one can get
$$
{W^{\prime\prime}\over{W}} = {8\pi v^{2}\over{W^{2}}}e^{2K}.
\eqno{(7a)}
$$
Again using (6c) and (6d) one can write
$$
-{W^{\prime\prime}\over{W}}+2U^{\prime\prime}+2U^{\prime}{W^{\prime}\over{W}}
= - {8\pi v^{2}\over{W^{2}}}e^{2K}.
\eqno{(7b)}
$$
Now adding (7a) and (7b) one gets
$$
U^{\prime\prime} + {U^{\prime}W^{\prime}\over{W}} = 0,
$$
which on integration yields
$$
U^{\prime} = {\alpha\over{W}},
\eqno{(8)}
$$
where $\alpha$ is an integration constant. Again by adding (6b) and
(6c) one can get
$$
K^{\prime\prime} + {K^{\prime}W^{\prime}\over{W}} = 0,
$$
which on integration yields
$$
K^{\prime} = {\beta\over{W}},
\eqno{(9)}
$$
where $\beta$ is another integration constant. Now putting (8) and (9)
in (6c) one can write
$$
K^{\prime\prime} + (A^{2} + B^{2} e^{2K}){K^{\prime}}^{2} = 0,
\eqno{(10)}
$$
where $A^{2} = {\alpha^{2}\over{\beta^{2}}}$ and $B^{2} = {4\pi
v^{2}\over{\beta^{2}}}$. For any arbitary value of $A$ and $B$ it is very
difficult to obtain a solution for (10) in closed form. However if one
defines a new coordinate 
$$
{u\over{u_{0}}} = exp(k),
$$
then one can find a solution for the complete spacetime which becomes
$$
ds^{2} = (u/u_{0})^{2(1-A)}dt^{2} - (u/u_{0})^{2A}dz^{2} -
(u/u_{0})^{2A(A-1)}exp[B^{2}(u/u_{0})^{2}]((1/u_{0}^{2})du^{2}
+ \beta^{2}d\theta^{2})
\eqno{(11a)}
$$ 
This solution is similar to that obtained earlier by Cohen and kaplan~\cite{CK}
for global string with bound matter at rest. The spacetime may have singularity 
at $u=0$ or at $u=\infty$ depending on the value of $A$ which is an
arbitary constant. 

For $A=1/2$ one can check that the spacetime admit the Lorentz boost and
spacetime (11a) becomes
$$
ds^{2} = (u/u_{0})(dt^{2} - dz^{2})
- (u/u_{0})^{-1/2}exp[B^{2}(u/u_{0})^{2}]((1/u_{0}^{2})du^{2}
+\beta^{2}d\theta^{2})
\eqno{(11b)}
$$
This is similar to that obtained by Cohen and Kaplan~\cite{CK} and also by
present authors ~\cite{Ban} and also has a singularity at finite distnace
outside the core. But one may notice that there is one difference between
the
metric (11a) and (11b) with the corresponding solutions given by Cohen
and Kaplan(CK). In CK solution, the exponential term in $g_{uu}$ and in 
$g_{\theta\theta}$ appears as the reciprocal of what we get here.

However, for $A^{2}=2$, one can get a closed form solution for the equation (10)
which is of the form
$$
e^{2K} = P + {2\over{B^{2}}} ln(r/r_{0}),
\eqno{(12)}
$$
where $r_{0} = 2/B = \beta/(\sqrt{\lambda} v)$, and $P$ is an arbitary
constant of integration. The other metric components in this case are
$$
e^{2U} = [P + {2\over{B^{2}}} ln(r/r_{0})]^{\sqrt{2}},
\eqno{(13)}
$$
$$
W = 2 B \beta (r/r_{0}) [P + {2\over{B^{2}}} ln(r/r_{0})].
\eqno{(14)}
$$
The line element can now be written as
$$
ds^{2} = [P + {2\over{B^{2}}} ln(r/r_{0})]^{(1-\sqrt{2})} (dt^{2} - dr^{2})
- [P + {2\over{B^{2}}} ln(r/r_{0})]^{\sqrt{2}} dz^{2}
$$
$$
 - 4 B^{2} \beta^{2}
(r/r_{0})^{2} [P + {2\over{B^{2}}} ln(r/r_{0})]^{(2-\sqrt{2})} d\theta^{2}.
\eqno{(15)}
$$
One can identify $r_{0}$ as the core radius $\delta$ of the string and
hence the above line element is valid for $r\ge r_{0}$. Now if 
$P$, which is an arbitary integration constant, is positive, then it can
beshown that there is
no singularity for $r\ge\delta$. To check this, one can calculate the
Kretschman scalar for the metric (1) which comes out as
$$
R^{\mu\nu\alpha\beta}R_{\mu\nu\alpha\beta} =
[\{(1-\sqrt{2})K^{\prime\prime}  + 2(1-\sqrt{2})^{2}K^{\prime 2}\}^{2} +
( \sqrt{2}K^{\prime\prime} + 4K^{\prime 2})^{2} + \{
{W^{\prime\prime}\over{W}} - \sqrt{2}K^{\prime\prime} -
$$
$$
4\sqrt{2}{K^{\prime}W^{\prime}\over{W}} + {W^{\prime 2}\over{W^{2}}} +
4K^{\prime 2} \}^{2}]e^{4(U-K)}.
\eqno{(16)}
$$
Now the different terms in the expression (16) look like
$$
K^{\prime} = {1\over{B^{2}r[P + {2\over{B^{2}}} ln(r/r_{0})]}},
\eqno{(17a)}
$$
$$
K^{\prime\prime} = -{P +{ 2\over{B^{2}}} + {2\over{B^{2}}} ln(r/r_{0})\over{
B^{2} r^{2}[P + {2\over{B^{2}}} ln(r/r_{0})]}^{2}},
\eqno{(17b)}
$$
$$
{W^{\prime}\over{W}} = {{P + {2\over{B^{2}}} + {2\over{B^{2}}}
ln(r/r_{0})}\over{ r[P + {2\over{B^{2}}} ln(r/r_{0})]}},
\eqno{(17c)}
$$
$$
{W^{\prime\prime}\over{W}} = {2\over{B^{2}r^{2}[P + {2\over{B^{2}}}
ln(r/r_{0})] }}.
\eqno{(17d)}
$$
>From expressions (17a) - (17d) and also from the
expressions for $e^{2U}$ and $e^{2K}$ one can check easily that the
Kretschmann scalar (16) approaches zero as $r\rightarrow \infty$ and for
all other value of $r>r_{0}$ the scalar is finite. Hence one can conclude
that the metric (15) for the spacetime outside the core of a global
string is nonsingular. This is perhaps the first example of a nonsingular
spacetime outside the core of a static global string. For obtaining such a
solution we have relaxed the condition of admitting a Lorentz boost.
Gregory
has shown~\cite{Greg1} that spacetime of global string 
should admit Lorentz boost if one has to demand elementary
flatness on the axis of the string. But as our spacetime is valid only
for outside the string core one can not demand elementary flatness on the
string axis. Hence Lorentz boost is not an essential condition for
spacetime outside the string core.
Another important thing is that for our spacetime (15) one can have
$P_{r}+P_{\theta}=0$ where $P_{r}$ and $P_{\theta}$ are the pressures 
along the radial and tangential direction respectively
and thus Gregory's conjecture, that this pecularity
of the energy momentum tensor leads to the singular nature of global
string spacetime, does not hold. It can also be checked that the energy
density $T_{t}^{t}$ goes to zero as $r\rightarrow \infty$ and also the
kretschmann scalar becomes zero for $r\rightarrow \infty$ as we have
mentioned earlier. Hence the solution is asymptotically well behaved 
which removes the unpleasant feature that a global string can not be
asymptotically wellbehaved ~\cite{Gib}.
It appears that all these properties are the result of relaxing the demand
of a Lorentz boost along the symmetry axis. To preserve the Lorentz boost,
One has to take $A^{2}=1/2$. For all other values of $A$, this symmetry is
lost.we can get
the solutions in a closed form only for $A^{2}=2$. For all other values of
$A$, a series solution in the form of $r=r(K)$ may be obtained (like the
one presented for $A^{2}=1/2$ in ref ~\cite{Ban}), which may or may not
have the singularity. In this work, we have thus been able to prove that
at least one nonsingular static global string solution can be obtained in
the absence of Lorentz boost along the z-axis.A more general study  
in this line will definitely be worthwhile.  
 
The authors wish to thank the members of Relativity and Cosmology
Research Centre, Jadavpur University, for the suggestions and comments.

\end{document}